# Unveiling the photoluminescence dynamics of gold nanoclusters with fluorescence correlation spectroscopy

Malavika Kayyil Veedu,[1] Julia Osmólska,[2] Agata Hajda,[2] Joanna Olesiak-Bańska,[2] Jérôme Wenger[1,*]

[1] Aix Marseille Univ, CNRS, Centrale Marseille, Institut Fresnel, AMUTech, 13013 Marseille, France

[2] Institute of Advanced Materials, Wroclaw University of Science and Technology, Wrocław, Poland

* Corresponding author: jerome.wenger@fresnel.fr

**Abstract:**
Gold nanoclusters (AuNCs) have captured significant interest for their photoluminescent properties; however, their rapid photodynamics remain elusive while probed by ensemble-averaging spectroscopy techniques. To address this challenge, we use fluorescence correlation spectroscopy (FCS) to uncover the photoluminescence dynamics of colloidal $Au_{18}(SG)_{14}$ nanoclusters. Our FCS analysis reveals the photoluminescence (PL) brightness per nanocluster, elucidating the impact of photoexcitation saturation and ligand interactions. Unlike DNA-encapsulated silver nanoclusters, the gold counterparts notably exhibit minimal blinking, with moderate amplitudes and 200 µs characteristic times. Our data also clearly reveal the occurrence of photon antibunching in the PL emission, showcasing the quantum nature of the PL process, with each AuNC acting as an individual quantum source. Using zero-mode waveguide nanoapertures, we achieve a 16-fold enhancement of the PL brightness of individual AuNCs. This constitutes an important enabling proof-of-concept for tailoring emission properties through nanophotonics. Overall, our study bridges the gap between ensemble-averaged techniques and single-molecule spectroscopy, offering new insights into AuNC photodynamics for biosensing and imaging applications.

**Keywords:** gold nanoclusters, fluorescence correlation spectroscopy FCS, fluorescence photodynamics, blinking, zero-mode waveguide nanoaperture

**Figure for Table of Contents:**

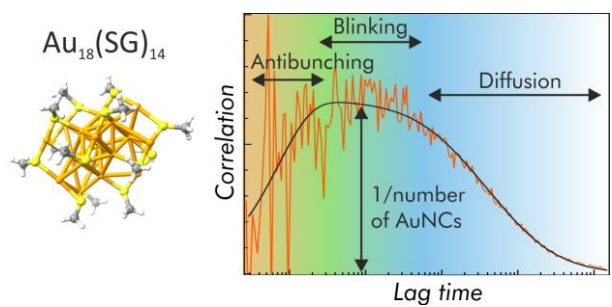

**Text for Table of Contents entry:** Atomically-precise gold nanoclusters, known for their unique photoluminescence, are rarely explored at the single nanocluster level. Our study using fluorescence correlation spectroscopy reveals key insights into their photoluminescence dynamics.

**Introduction**

Gold nanoclusters (AuNCs) represent a specialized class of atomically precise metal nanoparticles (NPs) with diameters typically below 2 nm. This small size attributes the nanoclusters with unique electronic and optical characteristics that diverge from larger nanoparticles, primarily due to their high surface-to-volume ratios, electron sharing mechanisms, and quantum confinement effects.[1,2] Existing at the interface between individual atoms and larger metal structures, gold nanoclusters offer crucial insights into the progressive changes in structure and properties that occur when scaling from atomic to nanoparticulate forms.[3] Noteworthy feature of AuNCs is the concept of a 'superatom', which refers to specific combinations of gold atoms, surface ligands, and charges that result in particularly stable nanocluster structures with closed electron structure.[4] Nanoclusters manifest molecular-like behavior, including discrete electronic structures and notably distinct optical properties, featuring luminescence ranging from visible to near infrared wavelengths, which originates from the S-Au-S-Au-S semi-ring states and the Au core states transitions.[5–7] As a result, AuNCs have gained significant attention for a multitude of applications, ranging from catalysis,[8,9] to optical devices,[10–12] biosensing,[13] and bioimaging.[14,15]

While gold nanoclusters are attracting a large interest, their photoluminescence (PL) properties remain primarily scrutinized through bulk ensemble-averaged spectroscopy. This approach, unfortunately, fails to unveil rapid photodynamic processes occurring below the millisecond time scale, leaving critical questions unanswered. Does the AuNC luminescence exhibit blinking behavior? If so, at what time scale? How does the brightness per nanocluster evolve with the excitation power? How many emitting centers are present on an AuNC? Ensemble measurements are unable to resolve the fast photodynamics processes, prompting the need to apply techniques from single-molecule fluorescence spectroscopy in order to find answers.[16,17] Among the different single-molecule techniques,[18] fluorescence correlation spectroscopy (FCS) is a versatile and powerful tool to investigate all the dynamic phenomena leading to a change in the fluorescence intensity, from photon antibunching to triplet state blinking and translational diffusion.[19,20]

Here, we apply FCS to unveil the photoluminescence dynamic properties of water-soluble $Au_{18}(SG)_{14}$ nanoclusters (SG – L-glutathione), with well-defined structure and good PL properties.[21,22] Our FCS data assess the PL brightness per individual nanocluster, allowing to properly report the transition into the saturation regime while increasing the excitation power. Our findings demonstrate that heavy water as solvent and pyridinedicarboxaldehyde (PDA) as ligand passivating shell significantly promote the PL brightness. Contrarily to DNA-encapsulated silver nanoclusters,[23] these gold nanoclusters exhibit moderate blinking, with amplitudes below 0.2 and characteristic times around 200 µs. Furthermore, our observations reveal the signature of photon antibunching in the nanocluster emission, highlighting the quantum nature of the PL process. This insight firmly establishes each AuNC as an individual quantum source featuring a single emitting center. Shifting away from the conventional confocal microscope system, we show that zero-mode waveguide nanoapertures of 110 nm milled in aluminum film can enhance the PL brightness of individual AuNCs by an impressive 16-fold factor. This proof-of-concept expands our understanding and paves the ways for using nanophotonics to tailor the AuNCs emission properties. Gaining deeper insights into the AuNC photodynamics beyond the capabilities of ensemble-averaging techniques represents a critical stride toward future advancements in deploying AuNCs for biosensing and imaging applications.

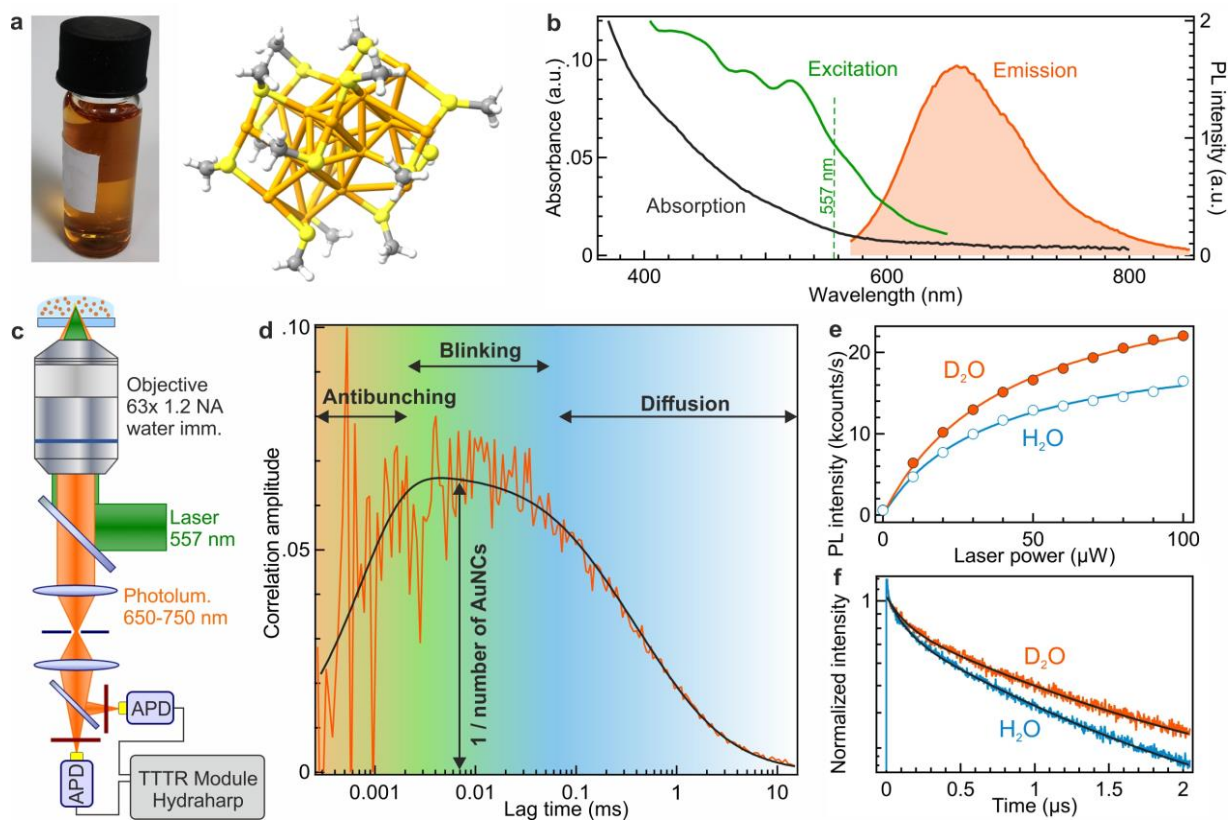

**Figure 1.** a) Vial containing $Au_{18}(SG)_{14}$ used for the measurements and schematic representation of the AuNC (orange: gold atoms, light yellow: sulfur, instead of glutathione, only simplified functional groups –SCH3 are represented). b) Absorption, excitation, and emission spectra of $Au_{18}(SG)_{14}$. We used an excitation laser of 557 nm for further experiments and APDs collecting photons in the range 650-750 nm. c) Scheme of the FCS setup used for our measurements. d) FCS correlation curve of $Au_{18}(SG)_{14}$ illustrating various photophysical processes occurring at different time scales. e) Comparison of the photoluminescent (PL) intensity and fluorescence decay trace of $Au_{18}(SG)_{14}$ dispersed in $H_2O$ and $D_2O$. (f) Comparison of the PL lifetime decays in $D_2O$ and $H_2O$.

## Results and Discussion

**Confocal measurements of $Au_{18}(SG)_{14}$**

The synthesized $Au_{18}(SG)_{14}$ nanoclusters (shown in Fig 1a) exhibit a quantum yield of 7% in pure water with an emission peak at 660 nm and a broad excitation spectrum in the blue-green region (as depicted in Fig 1b). Our primary objective is to gain a deeper understanding of their photophysical properties. To this end, we implement FCS to analyze the luminescence from a few nanoclusters in a well-defined confocal detection volume (Fig. 1c).[19] The FCS analysis allows to quantify various processes affecting the photoluminescence signal, as schematically depicted in Fig. 1d. The amplitude of the FCS correlation function is proportional to the inverse number of AuNCs present in the confocal detection volume, while the characteristic shape of the correlation function at different timescales provide information about different processes such as photon antibunching, blinking and Brownian diffusion as we will quantify later below.

A characteristic feature of AuNCs is the presence of the antibunching dip occurring at microsecond timescales in the FCS correlation function (Fig. 1d). For conventional organic fluorescent dyes with fluorescent lifetimes in the nanosecond range, this antibunching dip in the FCS function is rarely seen.[24,25] Instead, $Au_{18}(SG)_{14}$ NCs have a luminescence lifetime in the microsecond range, so that the antibunching dip can be clearly resolved in the FCS correlation. The physical origin of this dip relates to the quantum nature of the single photon emitted by a single AuNC.[24,26–28] The single photons are either transmitted or reflected at the 50/50 beamsplitter separating the avalanche photodiodes, but the two photodiodes cannot detect a correlated signal simultaneously at a lag time shorter than the typical PL lifetime. The presence of the antibunching dip in the $Au_{18}(SG)_{14}$ FCS data highlights the quantum nature of the PL emission and the fact that photons are emitted one by one for each nanocluster. Said differently, there is only one emitting center per nanocluster, the $Au_{18}(SG)_{14}$ investigated here do not carry multiple emitting centers per AuNC. While the antibunching time includes a dependence on the excitation rate (see Eq. (7) in the Methods section), the observation of the antibunching dip is not expected to depend on the choice of the excitation wavelength, alike the luminescence lifetime.

To improve the luminescence brightness of AuNCs, we replace water $H_2O$ by heavy water $D_2O$ as the solvent. The rationale behind this choice is motivated in the fact that O-D stretching has a lower vibrational frequency compared to O-H stretching, thereby reducing solvent-mediated non-radiative decays when using $D_2O$.[29,30] The outcomes presented in Fig 1e demonstrate a substantial 30% increase in PL intensity when $D_2O$ is used compared to $H_2O$. This luminescence intensity gain is accompanied with a PL lifetime increase as shown on Fig. 1f. In $H_2O$, the intensity-averaged lifetime is approximately 1190 ns, which increases to 1350 ns when $D_2O$ is used as buffer. This lifetime increase is reminiscent of the fluorescence lifetime gain observed with visible and near-infrared fluorescent dyes, and is consistent with a reduction of the nonradiative decay rate constant due to the reduced quenching losses with $D_2O$ instead of $H_2O$.[29,30] In the following, all the experiments use $D_2O$ instead of $H_2O$ to maximize the PL intensity.

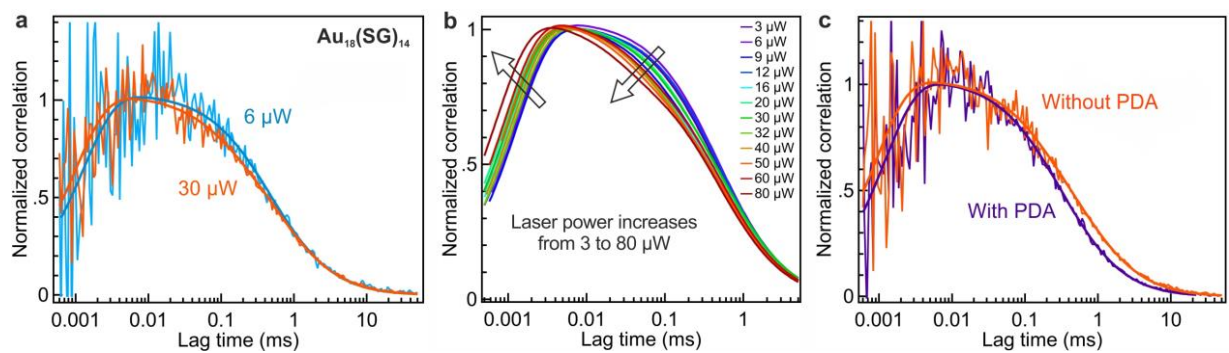

**Figure 2**. a) Normalized FCS correlation and numerical fits of Au$_{18}$(SG)$_{14}$ in D$_2$O at two different powers (6 µW and 30 µW). b) Comparison of the numerical fits of the normalized FCS correlation of Au$_{18}$(SG)$_{14}$ in D$_2$O with increasing the laser power from 3 µW to 80 µW. The trend in the numerical fits with increasing laser power is denoted by the arrow. c) Comparison of the normalized FCS correlations and fits of Au$_{18}$(SG)$_{14}$ in the presence and absence of PDA at 20 µW. The measurements of Au$_{18}$(SG)$_{14}$ without PDA are done in 98% v/v of D$_2$O and the measurements of Au$_{18}$(SG)$_{14}$ with PDA are done in a PBS buffer of pH 11. The excitation laser used is 557 nm.

**FCS data of Au$_{18}$(SG)$_{14}$**

Figure 2 presents FCS data for the different samples tested on the confocal setup. We aim to probe the evolution of the FCS data with the laser power (Fig. 2a,b) as well as the influence of the outer ligand surrounding the AuNC (Fig. 2c). It has been reported in Ref.[31] that the addition of pyridinedicarboxaldehyde (PDA) to glutathione-stabilized nanoclusters enhances the PL signal by forming a robust covalent imine bond between PDA and SG at pH 11. We have decided to apply a similar procedure to functionalize our Au$_{18}$(SG)$_{14}$ nanoclusters with PDA and compare their photophysical properties in the presence or absence of PDA using FCS.

Representative examples of correlation functions for Au$_{18}$(SG)$_{14}$ are shown on Fig. 2a at two different excitation powers and on Fig. 2c in presence or absence of the ligand exchange procedure involving PDA. By fitting the correlation functions using equation (1) (See Materials & Methods), we gain insights into the diffusion kinetics and photophysical characteristics of Au$_{18}$(SG)$_{14}$. The FCS data allows to determine the hydrodynamic radius $r_H$ of Au$_{18}$(SG)$_{14}$. For this, we use a separate calibration of the FCS diffusion time with Alexa Fluor 647 which is known to have a 0.7 nm hydrodynamic radius.[32] By using the equation $r_H/r_{Alexa647} = \tau_D/\tau_{D(Alexa647)}$, we obtain a hydrodynamic radius of 2.8 ± 0.5 nm for the Au$_{18}$(SG)$_{14}$.

Figure 2b highlight a discernible change in the shape of the fitted curve at different power levels. This is attributed to the dependence of photophysical parameters with the laser power and the occurrence of PL saturation due to the finite PL lifetime. As the laser power is increased (Fig. 2b) the antibunching dip seen at submicrosecond timescales becomes narrower as a consequence of the reduction of the antibunching time $\tau_A$ when the laser power is increased (see Eq. (7)). Similarly, the blinking time $\tau_{DS}$ gets shorter when the laser power is increased (as expected from Eq. (6)) and the fraction of molecules in the dark state $T_{DS}$ increases (see Eq. (5)). The different fit results are summarized in Fig. 3 and S3.

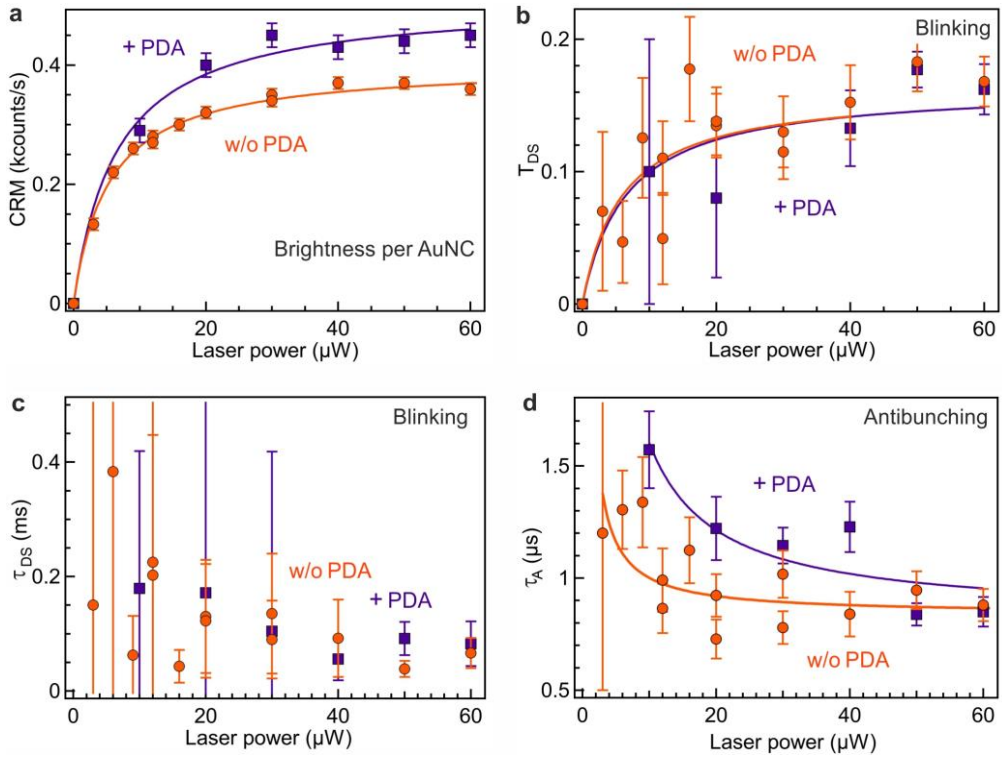

**Figure 3.** Comparison of different photophysical parameters of $Au_{18}(SG)_{14}$ obtained from the FCS data with increasing excitation laser power and in presence or absence of PDA. a) Fluorescence brightness per $Au_{18}(SG)_{14}$ with and without PDA. b) Dark state amplitude $T_{DS}$. c) Dark state blinking time $\tau_{DS}$. d) Antibunching time $\tau_A$.

**Photophysical parameters of $Au_{18}(SG)_{14}$**

Figure 3 summarizes our main results assessing the photophysical characteristics of $Au_{18}(SG)_{14}$ and their evolution as a function of the excitation power and the presence of PDA. The luminescence brightness per nanocluster CRM follows the power dependence described by Eq. (4) with a linear dependence as the laser power remains below the saturation power $P_{sat}$ of 5.5 µW, and a saturation when the laser power exceeds $P_{sat}$ (Fig. 3a). The presence of PDA induces a significant increase in the CRM of $Au_{18}(SG)_{14}$ nanoclusters. The saturation power in presence of PDA is also slightly raised up to 6.5 µW.

The observed blinking properties of $Au_{18}(SG)_{14}$ nanoclusters have a moderate amplitude $T_{DS}$ ranging from 0.04 to 0.2, contingent upon the power level (Fig. 3b), while the blinking time $\tau_{DS}$ remains on the order of 100 to 200 µs (Fig. 3c). The limited PL brightness per nanocluster (Fig 3a) leads to a relatively large noise on the determination of the blinking parameters. Figure S2 compares the fitting of the FCS data with and without taking into account the dark state blinking. The lower fitting residuals with the blinking model substantiates the occurrence of moderate dark state blinking in the $Au_{18}(SG)_{14}$ sample. The presence of PDA does not appear to modify the blinking properties. These properties come in stark contrast with the strong blinking properties observed among various DNA-encapsulated silver nanoclusters.[23] While these silver nanoclusters were reported to undergo photoinduced charge transfer between the silver core and the encapsulating DNA molecules,[23] our AuNCs remain largely free of photoinduced blinking.

In the microsecond range and below, all the FCS correlation functions exhibit a characteristic antibunching dip, with the respective antibunching time $\tau_A$ decreasing from from 1.5 µs to 0.8 µs as the laser power is increased (Fig. 3d). This evolution is expected from the model Eq. (7). The occurrence of antibunching confirms that the detected photons are coming from a single quantum emitter and not a collective emission of multiple emitting centers. The presence of PDA increases the antibunching time, indicating a similar increase in the PL lifetime confirmed by TCSPC measurements and the higher saturation power seen for PDA in Fig. 3a. These findings are in correlation to previous research on gold nanoclusters functionalized with PDA.[31] We can relate this PL lifetime increase to a reduction of the nonradiative decay rate constant in presence of PDA which further protects and stabilizes the AuNC.

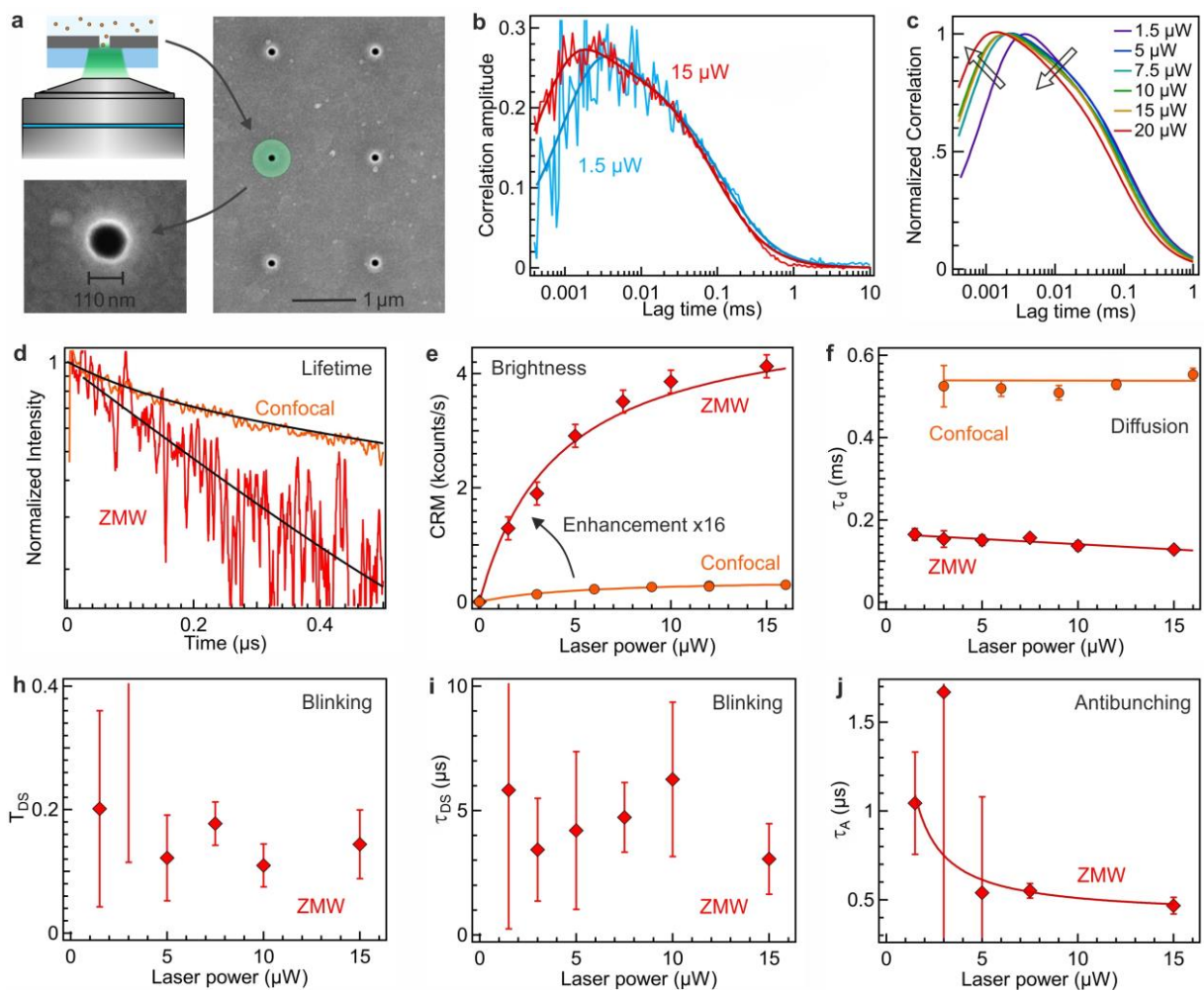

**Figure 4.** Enhancement of AuNC PL using zero-mode waveguide nanoapertures. a) Scanning electron microscope (SEM) image of aluminium ZMW aperture of 110 nm dimeter and its schematic representation at the microscope focus. b) FCS correlation and numerical fits of $Au_{18}(SG)_{14}$ in ZMW at two different powers (1.5 µW and 15 µW) of 557 nm excitation. c) Comparison of the numerical fits of the FCS correlation of $Au_{18}(SG)_{14}$ in a ZMW with increasing the laser power from 1.5 µW to 20 µW. The trends with increasing laser powers are denoted by the arrows. d) PL lifetime decay traces of $Au_{18}(SG)_{14}$ in confocal (diffraction-limited) and ZMW nanoaperture. e) Comparison of $Au_{18}(SG)_{14}$ brightness per nanocluster in the confocal and ZMW setup. f) Comparison of diffusion time of $Au_{18}(SG)_{14}$ in the confocal and in the ZMW. h-j) Evolution of different photophysical parameters of $Au_{18}(SG)_{14}$ in the ZMW obtained from the FCS data with increasing excitation laser power: h) dark state amplitude $T_{DS}$, i) dark state blinking time $\tau_{DS}$ and j) antibunching time $\tau_A$.

**PL enhancement using zero-mode waveguide nanoapertures**

To further improve the PL brightness of AuNCs and demonstrate that nanophotonics can be used to enhance their PL properties,[33,34] we use zero-mode waveguide (ZMW) nanoapertures. ZMWs are 110 nm diameter nanoapertures milled in an opaque aluminum film (Fig. 4a).[35–37] As the ZMW diameter is well below the optical wavelength, light cannot propagate through the aperture and thus the light intensity is confined at the bottom of the ZMW.[38] When a single ZMW is positioned at the microscope focus, it provides an effective way to confine light and locally overcome the diffraction limit. ZMWs have been reported to provide attoliter detection volumes together with 5 to 20× enhancement for the brightness of organic fluorescent dyes.[39,40] Here we demonstrate that the remarkable optical properties of ZMWs are fully applicable to AuNCs and can be used to further improve their brightness.

The correlation functions at different excitations powers are shown in Fig 4b,c, where we retrieve the characteristic photodynamic features seen in the confocal configuration (Fig. 1d & 2). In the ZMW, the PL lifetime is reduced as a consequence of the modification of the electromagnetic environment of the emitter[39,40] and becomes 1 µs (Fig 4d). The PL brightness per AuNC is significantly enhanced with the ZMW as compared to the conventional confocal FCS method (Fig 4e). At lower powers below saturation, the enhancement factor approaches 16×. As for organic fluorescent dyes, the brightness enhancement for AuNCs can be explained by a combination of local excitation intensity gain, quantum yield improvement and collection efficiency gain.[36,39,41] The saturation power is $\sim$ 4 µW, lower than that of confocal ($\sim$ 5.5 µW). While the confocal brightness remains around 400 counts/s at best (Fig. 3a), in the ZMW, PL brightness of 4000 counts/s are readily achieved, demonstrating the superior emission performance inside ZMWs.

The diffusion time $\tau_D$ in the ZMW amounts to 140 µs (Fig. 4f), and is lower than that of confocal measurements due to the smaller observation volume within the ZMW. As observed previously with fluorescent dyes, the reduction in the diffusion time $\tau_D$ is not proportional to the ratio of the illuminated areas, as the 3D shape of the ZMW nanoaperture requires additional time for the molecule to diffuse across the bottom of the ZMW.[39,40]

The blinking parameters $T_D$ and $\tau_{DS}$ in the ZMW are shown in Fig 4h,i. They display a moderate evolution with the laser power, which we relate to the occurrence of saturation at reduced powers and the larger statistical noise on the determination of the blinking dynamics. As for the confocal reference (Fig. 3b), we find that the blinking amplitude remains generally below 0.2, yet the blinking time appears to be accelerated to about 5 µs timescale. This apparent acceleration in the blinking dynamics can be related to the higher local excitation intensity inside the ZMW and to the presence of the metallic walls of the ZMW which promote nonradiative energy transfer to the metal. Lastly, we retrieve the characteristic antibunching dip in the FCS data, with the antibunching time summarized in Fig 4j. In agreement with the PL lifetime reduction in the ZMW, the antibunching time is also reduced is a similar manner. Altogether, the data in Fig. 4 demonstrate that ZMWs can be used as a high efficiency platform for different applications of AuNCs.

**Conclusions**

While AuNCs have attracted much attention as novel luminescent probes for biosensing and bioimaging, their photodynamics properties remained elusive as most studies focused on ensemble-averaged techniques. Here, we successfully implement fluorescence correlation spectroscopy to reveal the photoluminescence dynamic properties of $Au_{18}(SG)_{14}$ down to the submicrosecond timescale. Thanks to FCS, the PL brightness per nanocluster can be measured, enabling to determine the influence of photoexcitation saturation as well as the role of ligands surrounding the nanocluster. We have exemplified that the presence of PDA can significantly promote the PL brightness per single emitter and increase its PL lifetime. FCS simultaneously allows to record the blinking dynamics. Here we show that contrarily to DNA-encapsulated silver nanoclusters,[23] the gold nanoclusters show minimal blinking, with amplitudes below 0.2 and characteristic times around 200 µs in the confocal condition. This behavior enhances the potential applications of AuNCs in bioimaging. In addition, our FCS data show a clear antibunching peak at microsecond lag times. The observation of the antibunching phenomenon highlights the quantum nature of the PL process in the AuNC and importantly demonstrates that each AuNC behaves as a single quantum source. This means that each AuNC carries a single emitting center, or very much alike quantum dots, the whole AuNC participates to the PL emission process. Finally, we demonstrate that ZMW nanoapertures can be used to further enhance by 16-fold the PL brightness of individual AuNCs. This proof-of-principle demonstration importantly unlocks the use of nanophotonics to tailor the emission properties of AuNCs. Gaining extra insights on the AuNC photodynamics beyond those accessible to ensemble-averaging techniques is an important step for the future development of AuNCs in biosensing and imaging applications, especially under single-particle regime.

**Materials and Methods**

**Nanocluster synthesis protocol:** Gold nanoclusters were synthesized using protocol described in Ref.[34], however scale of synthesis was downscaled. Briefly, 100mg of L-glutathione (GSH), 0.4 ml of methanol and 0.4 ml of water were mixed in 25 mL round flask. After few minutes of dissolving GSH, $HAuCl_4 \cdot 3H_2O$ was added (200 µl, 0.6362 M) and mix for another 10 min. During this time, a gradual discoloration can be observed - from an intense yellow to an almost colorless. Then, solution was diluted with methanol to 10 ml. Next, slow reduction was performed using methanolic solution of $NaBH_3CN$ (1.5 ml, 220 mM). The solution was left for mixing for 1 hour. Gradual color change from yellow to orange to dark brown was observed. After 1 hour the precipitate was collected and washed with MeOH three times through centrifugal precipitation (10 mi/12000 rpm). Collected sediment was dissolved in small amount of water and washed three times with methanol/ethanol, each time centrifuged (10 min/12000 rpm) to remove the remaining reaction precursors.

**Sample Preparation:** The gold nanocluster sample ($Au_{18}SG_{14}$) is diluted in deuterium oxide ($D_2O$). $D_2O$ is purchased from Sigma-Aldrich and used as received. The v/v ratio of nanocluster with $D_2O$ is 98% for confocal and 50% for ZMW measurements as for ZMWs we are limited by the minimal µM concentration required to perform FCS experiments and cannot dilute more than twice the stock AuNC solution. Before use, the ZMW nanoapertures are cleaned with Milli-Q water followed by rinsing with 97% ethanol. Then the ZMW sample is exposed to UV for 5 min to clean any organic impurities. 2,6-pyridine carboxaldehyde is purchased from Sigma-Aldrich. Synthesis of PDA-coated AuNCs is done following a reported protocol.[31] The confocal experiments with this PDA-coated sample are done in PBS buffer at pH 11 as we found that this system is unstable in $D_2O$ at pH 7.

**FCS setup:** All the fluorescence measurements are done in a home-built confocal microscope setup. $Au_{18}NC$ samples are excited at 557 nm by an iChrome-TVIS laser (Toptica GmbH, pulse duration ~3 ps). The repetition rate of the laser is 40 MHz. A multiband dichroic mirror (ZT 405/488/561/640rpc, Chroma) reflects the laser towards the microscope, and a Zeiss C-Apochromat 63×, 1.2 NA water immersion objective lens is used to focus the excitation light. The same objective lens collects the PL signal in an epifluorescence configuration. The PL beam then passes through the same multiband dichroic mirror. To block the laser back reflection an emission filter (ZET405/488/565/640mv2, Chroma) is used. The fluorescence signal is focused onto an 80 µm pinhole. Two avalanche photodiodes APDs (Perkin Elmer SPCM-AQR-13) separated by a 50/50 beam-splitter in a Hanbury-Brown-Twiss configuration record the emitted photons in the 650-750 nm spectral range. The photodiode outputs are connected to a time-correlated single photon counting (TCSPC) module (HydraHarp 400, Picoquant). The integration time for each FCS experiment was set to 40 minutes for the confocal experiments, and 5 minutes for the ZMWs.

**FCS Analysis:** FCS traces are obtained from the cross-correlation of the fluorescence intensity time trace from two APDs. We time-gated and discarded photons within the 0-2 ns range out of the 25 ns pulse interval to reduce the background due to laser-induced backscattering and Rayleigh scattering (Fig. S1). We used the following three-dimensional Brownian diffusion model with additional terms for blinking and anti-bunching effects to fit the FCS curves:[19]

$$G(t) = \frac{1}{N}\left(1 - \frac{1}{N_{emi}}e^{\left(\frac{-t}{\tau_A}\right)}\right)\left(1 + \frac{T_{DS}}{(1-T_{DS})}e^{\left(\frac{-t}{\tau_{DS}}\right)}\right)\left(1 + \frac{t}{\tau_D}\right)^{-1}\left(1 + \frac{t}{\kappa^2\tau_D}\right)^{-0.5} \quad (1)$$

Where G(t) is the cross-correlation function at time t, N is the total number of AuNCs in the observation volume, $N_{emi}$ is the number of emitting species per AuNC, $\tau_D$ is the mean diffusion time, $T_{DS}$ is the fraction of emitters in the dark state, $\tau_{DS}$ is the dark state blinking time, $\tau_A$ is the antibunching time and κ corresponds to the aspect ratio of the axial to the transversal dimension of the detection volume. Our different FCS data converge toward similar values $N_{emi}$ around one, so for consistency in the analysis, we decided to keep the $N_{emi}$ parameter at 1 for all the analysis. We have taken κ as 5 for confocal and κ as 1 for ZMW nanoaperture based on our past results as this fits well with our experimental FCS data. To extract different photophysical parameters from equation (1) we fit the experimental data from .01 μs to 100 ms. From the fitted FCS curve we extract the value of N, which is then corrected for the background (B) using equation (2) to determine the background corrected value for $N_{corr}$

$$N_{corr} = N\left(1 - \frac{B}{F}\right)^2 \qquad (2)$$

Here B is the measured background and F is the total PL intensity. $N_{corr}$ is defined as the mean number of detected fluorescent molecules in the observation volume averaged over the duration of the experiment. We also calculate the brightness per nanocluster CRM using

$$CRM = \frac{1}{N_{corr}}(F - B) \qquad (3)$$

The evolution of the brightness per nanocluster CRM is given by [18]

$$CRM = A\frac{P}{1+\frac{P}{P_{sat}}} \qquad (4)$$

Here $A = \eta\phi\sigma\rho$ is the product of collection efficiency η, the quantum yield ϕ, the excitation cross-section σ and a constant proportionality parameter ρ to accommodate for the different units while expressing the excitation power P in microwatts. $P_{sat}$ is the saturation power of the AuNC.

The dark state amplitude $T_{DS}$ follows a similar power dependence given by [42–44]

$$T_{DS} = \alpha\frac{P}{1+\frac{P}{P_{sat}}} \qquad (5)$$

where $\alpha = \frac{\sigma k_{isc} \rho}{k_{ph} k_{tot}}$, $k_{isc}$ and $k_{ph}$ are the rate constants of inter-system crossing and dark state de-excitation respectively and $k_{tot}$ is the total de-excitation rate constant.

The blinking time $\tau_{DS}$ is related to the different decay rate constants as [42–44]

$$\frac{1}{\tau_{DS}} = k_{ph} + \frac{\sigma k_{isc} \rho}{k_{tot}}\frac{P}{1+\frac{P}{P_{sat}}} \qquad (6)$$

Lastly, the antibunching time $\tau_A$ is given by [26]

$$1/\tau_A = k_{tot} + \sigma \rho P \qquad (7)$$

**Zero mode waveguide fabrication:** The focused ion beam (FIB) technique is used to fabricate zero-mode waveguide nanoapertures into an aluminum film of thickness 100 nm deposited on a glass coverslip. The deposition of aluminum on a glass coverslip is carried out by electron beam evaporation (Bühler Syrus Pro 710). The chamber pressure is set at $5 \times 10^{-7}$ mbar and the aluminum is deposited at the rate of 10 nm/s to achieve the optimum performance of our ZMW. A focused gallium beam (10 pA current and 30 kV voltage) with a resolution of 10 nm is then used to mill ZMW nanoapertures of diameter 110 nm into the aluminum film. A 12 nm-thick silica layer is deposited with plasma-enhanced chemical vapour protection (Oxford Instruments PlasmaPro NGP80) to protect the aluminum sample.

**Supporting Information**

TCSPC decays of background, Comparison of FCS fitting with and without the dark state blinking term, Supplementary FCS fit results

**Acknowledgments**

This project has received funding from the European Research Executive Agency (REA) under the Marie Skłodowska-Curie Actions doctoral network program (grant agreement No 101072818) and Sonata Bis project from the National Science Center in Poland (2019/34/E/ST5/00276).

**Conflict of Interest**

The authors declare no conflict of interest.

**Data Availability Statement**

The data that support the findings of this study data are available from the corresponding author upon request.

**Supporting Information for**

**Unveiling the photoluminescence dynamics of gold nanoclusters with fluorescence correlation spectroscopy**

Malavika Kayyil Veedu,[1] Julia Osmólska,[2] Agata Hajda,[2] Joanna Olesiak-Bańska,[2] Jérôme Wenger[1,*]

[1] *Aix Marseille Univ, CNRS, Centrale Marseille, Institut Fresnel, AMUTech, 13013 Marseille, France*

[2] *Institute of Advanced Materials, Wroclaw University of Science and Technology, Wrocław, Poland*

*\* Corresponding author: jerome.wenger@fresnel.fr*

**Contents:**

S1.  **TCSPC decays of background**

S2.  **Comparison of FCS fitting with and without the dark state blinking term**

S3.  **Supplementary FCS fit results**

**S1. TCSPC decays of background**

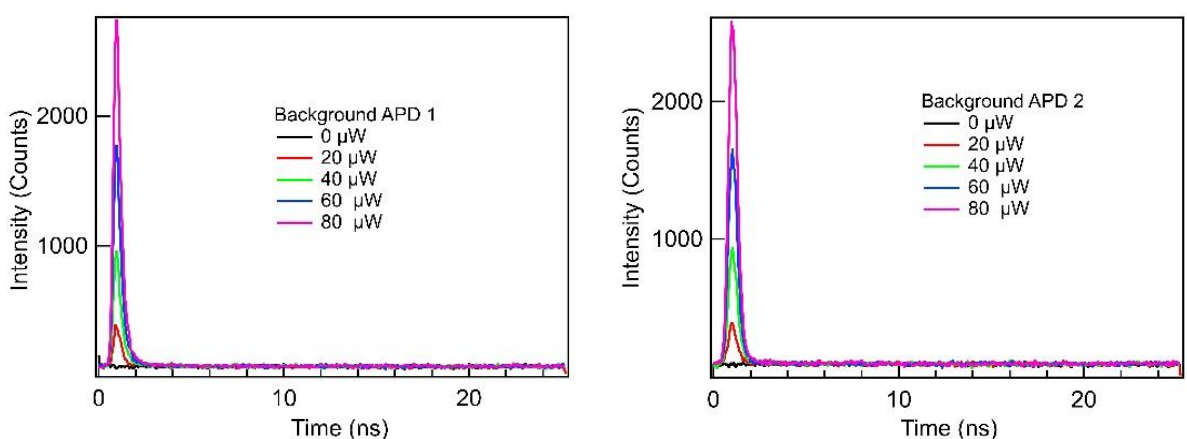

**Figure S1.** TCSPC measurement of $D_2O$ background at different powers. Here the fluorescence intensity counts in the first 2 ns are mainly due to laser-induced backscattering and Raman scattering. Hence we time gate intensity counts at the first 2 ns to remove the scattering peak and reduce the background.

**S2. Comparison of FCS fitting with and without the dark state blinking term**

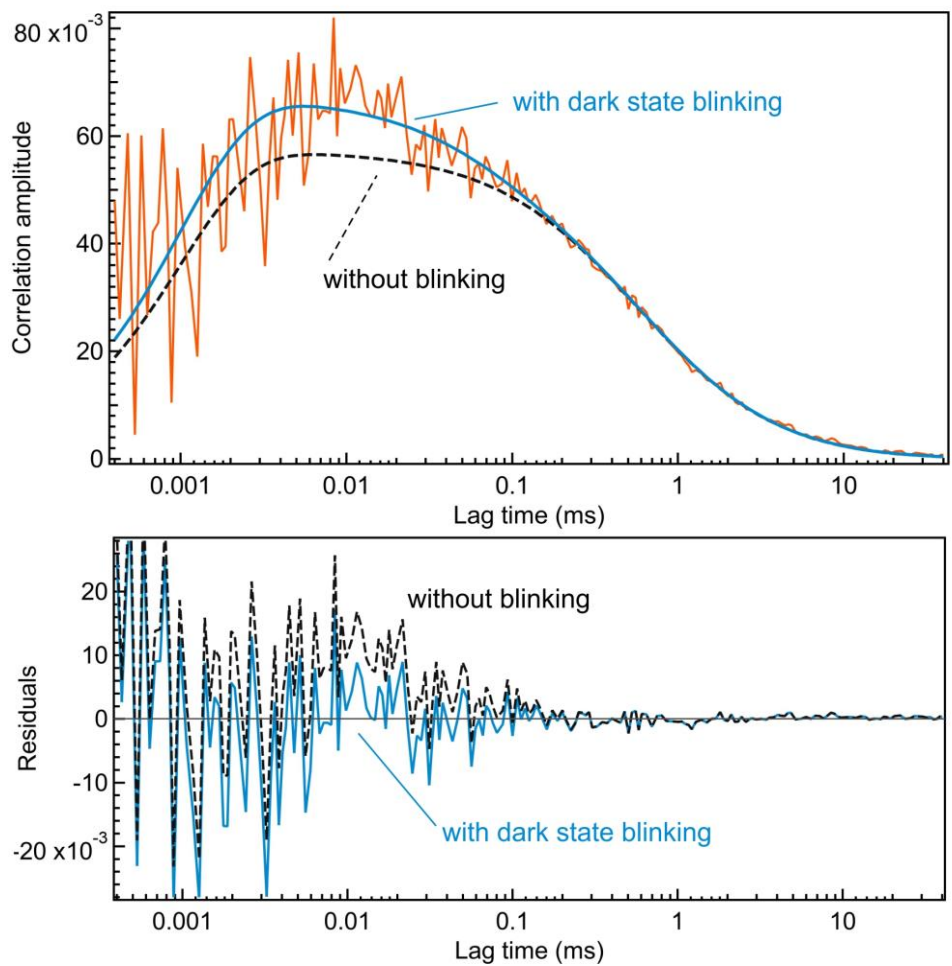

**Figure S2.** Comparison of FCS numerical fits with (solid blue line) and without (dashed gray line) including the term $\left(1 + \frac{T_{DS}}{(1-T_{DS})} e^{\left(\frac{-t}{\tau_{DS}}\right)}\right)$ accounting for the dark state blinking. The lower graph shows the fit residuals. The excitation power is 30 µW.

## S3. Supplementary FCS fit results

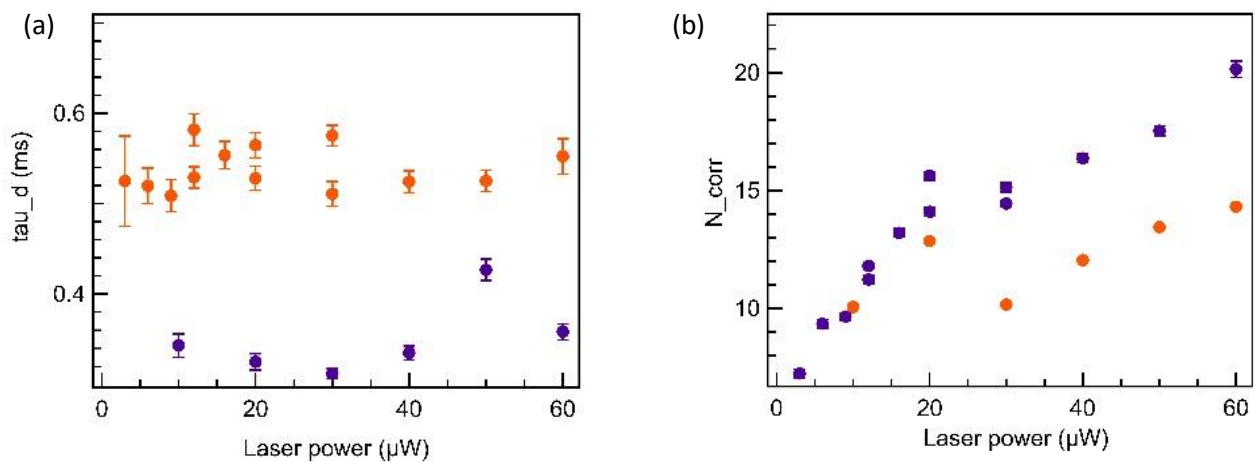

**Figure S3.** a) Diffusion time of Au$_{18}$(SG)$_{14}$ with/without PDA at different powers. b) Variation in the number of AuNC in the confocal detection volume N$_{corr}$ with/without PDA with power.